\documentclass[preprint2]{aastex}

\newcommand{\kt}{\ensuremath{k_{\rm{B}}T}}
\newcommand{\lx}{\ensuremath{L_{\rm{X}}}}

\newcommand{\nh}{\ensuremath{N_{\rm H}}}

\slugcomment{}
\shorttitle{Iron Fluorescent Line from YSOs in Orion Nebula}
\shortauthors{Tsujimoto et al.}

\begin{document}
\title{Iron Fluorescent Line Emission from Young Stellar Objects in the Orion Nebula}
\author{M.~Tsujimoto\altaffilmark{1}, E.~D.~Feigelson\altaffilmark{1},
N.~Grosso\altaffilmark{2}, G.~Micela\altaffilmark{3}, Y.~Tsuboi\altaffilmark{4},
F.~Favata\altaffilmark{5}, H.~Shang\altaffilmark{6}, J.~H.~Kastner\altaffilmark{7}}
\altaffiltext{1}{Department of Astronomy \& Astrophysics, The Pennsylvania State
University, 525 Davey Laboratory, University Park, PA 16802}
\altaffiltext{2}{Laboratoire d'Astrophysique de Grenoble, 414 rue de la Piscine,
Universit\'{e} Joseph-Fourier BP 53, 38041 Grenoble cedex 9, France}
\altaffiltext{3}{INAF, Osservatorio Astronomico di Palermo, Piazza del Parlamento 1,
90134 Palermo, Italy}
\altaffiltext{4}{Department of Physics, Chuo University, Kasuga 1-13-27, Bunkyo-ku,
Tokyo 112-8551, Japan}
\altaffiltext{5}{Astrophysics Division, Research and Science Support Department of ESA,
ESTEC, Postbus 299, 2200 AG Noordwijk, The Netherlands}
\altaffiltext{6}{Institute of Astronomy and Astrophysics, Academia Sinica, Taipei 106,
Taiwan, R.O.C.}
\altaffiltext{7}{Center for Imaging Science, Rochester Institute of Technology,
Rochester, NY 14623}

\begin{abstract}
 We present the result of a systematic search for the iron K$\alpha$ fluorescent line
 at $\sim$~6.4~keV among 1616 X-ray sources detected by ultra-deep \textit{Chandra}
 observations of the Orion Nebula Cluster and the obscured Orion Molecular Cloud 1
 population as part of the \textit{Chandra} Orion Ultra-deep Project (COUP). Seven
 sources are identified to have an excess emission at $\sim$~6.4~keV among 127 control
 sample sources with significant counts in the 6.0--9.0~keV band. These seven sources
 are young stellar objects (YSOs) characterized by intense flare-like flux variations,
 thermal spectra, and near-infrared (NIR) counterparts. The observed equivalent widths
 of the line cannot be attributed to the fluorescence by interstellar or circumstellar
 matter along the line of sight. The X-ray spectral fits and NIR colors of the 6.4~keV
 sources show that these sources have X-ray absorption of
 $\gtrsim$~1$\times$10$^{22}$~cm$^{-2}$ and NIR excess emission, which is not expected
 when the fluorescence occurs at the stellar photosphere.  We therefore conclude that
 the iron fluorescent line of YSOs arises from reflection off of circumstellar disks,
 which are irradiated by the hard X-ray continuum emission of magnetic reconnection
 flares.
\end{abstract}

\section{INTRODUCTION}
Contrary to most X-ray emission lines that convey information of ``hot'' celestial
objects, fluorescent emission carries information of ``cold'' matter in the
vicinity of bright X-ray sources. For a wide range of incident X-ray spectra, iron is the
most visible element in the K fluorescent lines from the photoionized matter when its
cosmic abundance and fluorescence yield are taken into account \citep{george91}. The
K$\alpha$ fluorescent line from neutral to low-ionized irons at $\sim$~6.4~keV can be
distinguished from K$\alpha$ lines from highly ionized irons ($\sim$~6.7~keV by
Fe~$_{\rm{XXV}}$ and $\sim$~6.9~keV by Fe~$_{\rm{XXVI}}$) using X-ray charge coupled
device (CCD) spectroscopy with a resolution of $\sim$~150~eV at $\sim$~6~keV. The line can
also penetrate through a large column density up to a few times 10$^{24}$~cm$^{-2}$,
which gives the opacity $\tau = 1$ by photoelectric absorption. All these features
make iron K$\alpha$ fluorescence a unique observational tool to investigate the
environment where the continuum X-ray sources reside.

The fluorescent line is reported in a variety of X-ray emitters. The two most widely known
classes are X-ray binaries and active galactic nuclei (AGNs). The detections of the line
also extend to giant molecular clouds (e.g., Sgr B2; \citealt{koyama96}), massive stars
(e.g., $\eta$ Car; \citealt{corcoran98}), supernova remnants (e.g., RCW\,86;
\citealt{vink97}), and unidentified sources in the Galactic center \citep{park04}. In
$\eta$ Car, \citet{corcoran98} attributed the line to the fluorescence from photoionized
circumstellar matter along the line of sight.

The best studied case of this emission is the Sun. The 6.4~keV line was first discovered
during solar flares \citep{neupert67}. Following studies revealed that the
photoionization by flare X-rays is more likely to produce this line than collision
ionization by accelerated particles \citep{doschek72}. \citet{parmar84} made a systematic
survey of the 6.4~keV emission among $\sim$~600 solar flares detected by \textit{Solar
Maximum Mission}. Their result was consistent with the calculation by \citet{bai79}, who
made a model assuming the 6.4~keV line is iron fluorescence at the solar photosphere.

Recently, \citet{imanishi01} detected the line emission from a class~I
protostar in $\rho$ Ophiuchi dark cloud at $\sim$~145~pc \citep{dezeeuw99} with
\textit{Chandra}. The protostar YLW\,16A produced a huge flare during the observation
reaching a luminosity of 1.4$\times$10$^{32}$~ergs~s$^{-1}$ (0.5--9.0~keV), during which
a 6.4~keV line emerged in its X-ray spectrum with an equivalent width (EW) of
$\lesssim$~180~eV. They interpreted that the emission is from a face-on protostellar disk
irradiated by the intense X-rays from a magnetic reconnection event. The previous
\textit{ASCA} detection of a $\sim$~6.2~keV emission line from an embedded far-infrared
source in the R CrA cluster \citep{koyama94} might be another example. These detections
point out the potential importance of this X-ray line to diagnose deeply embedded
protostars and their surroundings, although the geometry of the reflection for the
fluorescence is not certain with only a few detections. Continuum sources are presumably
flare X-rays, but reflectors can be interstellar matter along the line of sight, disks,
or photosphere. Systematic studies with more samples are mandatory to elucidate the
general picture of this faint emission.

\medskip
This paper has two purposes: (1) to establish the presence of the iron fluorescent line
from young stellar objects (YSOs) with a larger number of detections and (2) to examine
the geometry of reflection by comparing the 6.4~keV sample to a control sample not
showing the line. We utilize the most sensitive X-ray observation ever obtained of a
star-forming region, the \textit{Chandra} Orion Ultra-deep Project (COUP), to conduct a
systematic search for sources with an excess emission at $\sim$~6.4~keV. We report the
detections of seven YSOs with the iron fluorescent line and present the interpretation
that these fluorescent lines occur following irradiation of disks by intense flare
X-rays.

\section{OBSERVATION \& DATA REDUCTION}
The COUP observations were carried out in 2003 January for 13 consecutive days with a
total exposure time of $\sim$~838~ks. ACIS (Advanced CCD Imaging Spectrometer;
\citealt{garmire03}) on-board \textit{Chandra} \citep{weisskopf02} was used to take an
X-ray image covering a 17\arcmin$\times$17\arcmin\ region in the Orion Nebula Cluster
(ONC) at a distance of $\sim$~450~pc \citep{genzel89}. The image was centered at
R.\,A.\,$=$\,5$^{\rm h}$35$^{\rm m}$17$^{\rm s}$ and
Decl.\,$=$\,--5\arcdeg23\arcmin40\arcsec\ (J2000.0) to cover the entire ONC and its
vicinity, including the Orion Molecular Cloud 1 cores. The data were taken with  the
very faint telemetry mode and the timed exposure CCD operation with a frame time of 3.2~s.

Details of the data reduction are described in \citet{getman04}. Briefly, after removing
non-X-ray events and correcting for charge transfer inefficiency, a source search was
performed to detect 1616 X-ray sources. They were matched with various optical and
near-infrared (NIR) catalogs for identification. X-ray events of each source were
accumulated from an $\sim$~87\% encircled energy polygon at the source
position. Consistent imaging, spectral, and timing analyses were conducted for all
sources. In the spectral analysis, 0.5--8.0~keV spectra were fit by a thin-thermal
plasma model (mekal; \citealt{mewe85,liedahl95}) with interstellar absorption (wabs),
varying X-ray luminosity (\lx), plasma temperature (\kt), and equivalent hydrogen column
density (\nh) values. When a large residual was found, another temperature component was
added to improve the fits.

\section{SEARCH FOR FLUORESCENT LINE EMISSION}
\subsection{Control Sample and Event Extraction}
Among the 1616 COUP sources, we first define a control sample to search for the 6.4~keV
emission, which is restricted to sufficiently bright and spectrally hard sources. We set
two criteria of $>$ 100 photons and signal-to-noise ratio of $>$ 20 in the 6.0--9.0~keV
band to obtain a sample consisting of 127 sources. Here, we used a maximum energy of
9.0~keV rather than 8.0~keV used by \citet{getman04} to better determine the continuum
level beyond the iron K$\alpha$ lines at 6.4--6.9~keV.

Forty nine of these 127 sources have a ``p'' flag in the COUP source list
\citep{getman04}, indicating that they are possibly piled-up. We developed a method for
pile-up mitigation based on photon extraction in an annulus around the pile-up core of
the point spread function\footnote{More details are given by \citet{getman04} and at
http://www.astro.psu.edu/users/tsujimot/arfcorr.html.}. We accumulate X-ray events from
an $\sim$~87\% encircled energy polygon with the central circular core removed. We
derive an energy-dependent correction to the auxiliary response file for the annular
region based on simulations using ChaRT and MARX software packages\footnote{See
http://asc.harvard.edu/chart/ and http://space.mit.edu/CXC/MARX/.}. The correction
function is a low-order polynomial that does not introduce artificial features in the
corrected effective area function. Unlike the method by \citet{davis01} which is
restricted to mildly piled-up sources, our procedure is successfully applied to both
mildly and severely piled-up sources. Events extracted from the annular regions were
confirmed to meet the criteria for the control sample.

\subsection{Spectral Fitting}
For the 127 hard and bright COUP sources, 44 were successfully fit with
one-temperature thermal models and 83 were fit with two-temperature models, following
the procedure given by \citet{getman04}. We made fits in the 2.0--9.0~keV band with \nh\
fixed to the values tabulated by \citet{getman04}. The iron metallicity value was
treated as a free parameter, while the metallicity of other elements was left fixed to
0.3 solar.

We added onto the best-fit plasma model a Gaussian line component at 6.40~keV with the
intrinsic width fixed to 0 and a free normalization ($N_{\rm{K\alpha}}$). Here, 6.40~keV
is the weighted mean of two unresolved K$\alpha$ fluorescent lines from neutral irons at
6.391~keV and 6.404~keV (K$\alpha_{1}$ and K$\alpha_{2}$; \citealt{george91}). The
energy shift of the line center due to ions at different ionization stages makes a
negligible contribution with $\lesssim$~1\% for neutral to low-ionized irons
(Fe~$_{\rm{I}}$--Fe~$_{\rm{IX}}$; \citealt{house69}). The intrinsic width of these lines
($\sim$~0.4~eV and $\sim$~3.5~eV for thermal and quantum broadenings;
\citealt{george91}) and the width by the blending ($\sim$~12~eV) are both negligible
compared to the energy resolution of the detector ($\sim$~150~eV). We derive
$N_{\rm{K\alpha}}$, the EW of the line ($EW_{\rm{K\alpha}}$), and the null hypothesis
probability of the F test ($P_{\rm{F}}$) to examine whether an additional model is
statistically justified. With a careful inspection of the result, we recognized spectra
to have a 6.4~keV feature when they meet (1) the 90\% lower limit of $N_{\rm{K\alpha}}$
exceeds 0 and (2) $P_{\rm{F}}$ is less than 1\%. This suppresses the number of false
positives to be $\sim$~1. We further removed manually a few suspicious sources that
satisfied these criteria but did not visually show convincing fluorescent line emission;
these sources are pile-up sources that have distorted spectra even in reduced event
accumulation regions.

\begin{figure}
 \figurenum{1}
 \epsscale{1.0}
 \plotone{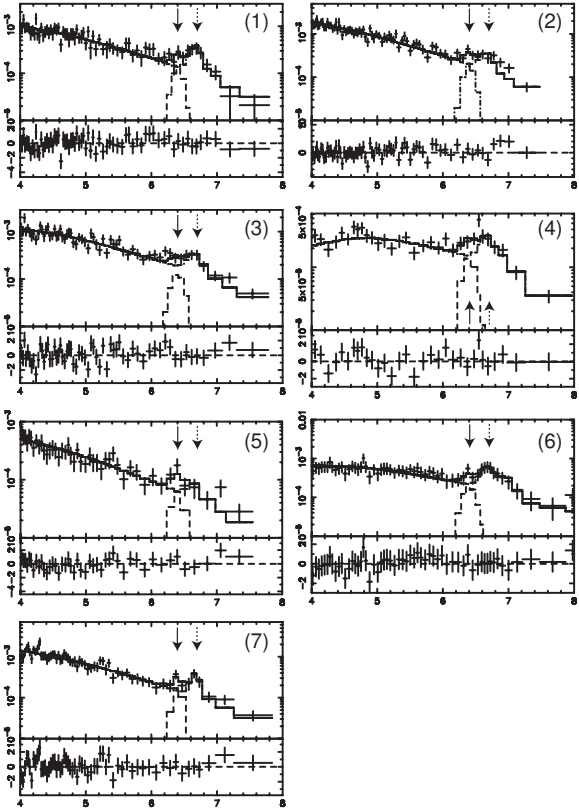}
 \caption{Spectra and best-fit models of 6.4~keV sources. The upper panels show the
 spectra (\textit{pluses}) and the best-fit models (\textit{solid steps}) with a
 Gaussian component shown by dashed steps. The 6.4 and 6.7 keV lines are indicated by
 solid and broken arrows, respectively. The abscissa is the energy from 4.0--8.0~keV,
 while the ordinate is the spectral intensity in the unit of
 cnts~s$^{-1}$~keV$^{-1}$. The lower panels show the residual to the fit, where the
 ordinate shows $\chi$ values of each bin.}\label{fg:f1}
\end{figure}

This procedure produces seven positive fluorescent line detections out of the 127
sources. Table~\ref{tb:t1} lists the name, the COUP sequence number, and the best-fit
values both of the Gaussian and thermal spectral components of the seven sources, while
the spectra of these sources are displayed in Figures~\ref{fg:f1}.

\begin{rotate}
\begin{deluxetable}{lcrrccccccc}
 \tabletypesize{\scriptsize}
 \tablecaption{X-ray Fitting Result of 6.4~keV Sources\label{tb:t1}}
 \tablecolumns{11}
 \tablewidth{0pt}
 \tablehead{
 \colhead{No.\tablenotemark{a}} &
 \colhead{Name} &
 \colhead{COUP \tablenotemark{b}} &
 \colhead{Cnts\tablenotemark{c}} &
 \colhead{$N_{\rm{K\alpha}}$\tablenotemark{de}} &
 \colhead{$EW_{\rm{K\alpha}}$\tablenotemark{e}} &
 \colhead{$P_{\rm{F}}$\tablenotemark{e}} &
 \colhead{\nh\tablenotemark{f}} &
 \colhead{$k_{\rm{B}}T$\tablenotemark{df}} &
 \colhead{$L_{\rm{X}}$\tablenotemark{df}} &
 \colhead{$L_{\rm{X,peak}}$\tablenotemark{g}} \\
 \colhead{} &
 \colhead{} &
 \colhead{Seq.\,No.} &
 \colhead{} &
 \colhead{(10$^{-7}$ cm$^{-2}$~s$^{-1}$)} &
 \colhead{(eV)} &
 \colhead{(10$^{-3}$)} &
 \colhead{(10$^{22}$ cm$^{-2}$)} &
 \colhead{(keV)} &
 \colhead{(10$^{30}$ ergs s$^{-1}$)} &
 \colhead{(10$^{31}$ ergs s$^{-1}$)}
 }
 \startdata
 1 & 053509.2$-$053058 & 331 & 268 & 2.6 (0.1--4.6) & 126 & 5.6 & \phn1.58 & 4.6 (4.3--5.6) & 3.4 (2.7--3.9) & 17\\
 2 & 053513.6$-$051954 & 561 & 332 & 3.6 (1.1--7.3) & 130 & 0.8 & \phn1.45  & 2.8 (2.6--2.9) & 5.6 (5.8--6.1) & 4.4\\
 3 & 053514.3$-$052232 & 621 & 304 & 2.5 (0.5--3.9) & 111 & 6.2 & \phn5.62 & 3.2 (2.9--3.4) & 4.1 (3.4--4.8) & 1.6\\
 4 & 053514.6$-$052211 & 647 & 272 & 4.1 (1.9--6.2) & 135 & 5.7 & 30.9\phn & 5.7 (4.7--8.1) & 3.3 (2.9--4.1) & 5.7\\
 5 & 053514.6$-$052301 & 649 & 107 & 1.2 (0.3--2.2) & 268 & 6.4 & \phn0.93 & 3.0 (2.6--3.1) & 1.4 (1.0--1.7) & 2.1\\
 6$^{\dagger}$ & 053519.3$-$052542 & 1030 & 461 & 17 (2.5--17) & 111 & 5.9 & 11.5\phn & 9.4 (6.4--14)\phd & 11\phd (9.3--12) & 45 \\
 7 & 053519.6$-$051326 & 1040 & 266 & 4.3 (1.5--6.6) & 122 & 6.0 & \phn1.91 & 3.3 (3.0--3.5) & 6.7 (6.2--7.2) & 40\\
 \enddata
 \tablenotetext{a}{Piled-up source for which we applied our pile-up mitigation method
 is marked with a dagger.}
 \tablenotetext{b}{Sequence number of the COUP sources \citep{getman04}.}
 \tablenotetext{c}{Counts in the 6.0--9.0~keV range. For the pile-up source, counts in the annulus region is given.}
 \tablenotetext{d}{The 90\% uncertainty is given in parentheses.}
 \tablenotetext{e}{The fitting result of the additional 6.4 keV Gaussian to the best-fit
 thin-thermal plasma model.}
 \tablenotetext{f}{The fitting result of the thin-thermal plasma model. For
 two-temperature spectra, the values of the higher temperature component are given.}
 \tablenotetext{g}{The peak X-ray luminosity during the flares. The values were derived
 from the maximam and average count rates among segmented blocks assuming no
 spectral change.}
\end{deluxetable}
\end{rotate}

\subsection{Light Curves}
The light curves were constructed for all the control sample sources using events in the
2.0--9.0~keV band and were segmented into pieces with a constant flux using a Bayesian
block algorithm \citep{scargle98}. The details on the application of algorithm to the
COUP data are given in \citep{getman04}.

\begin{figure}
 \figurenum{2}
 \epsscale{1.0}
 \plotone{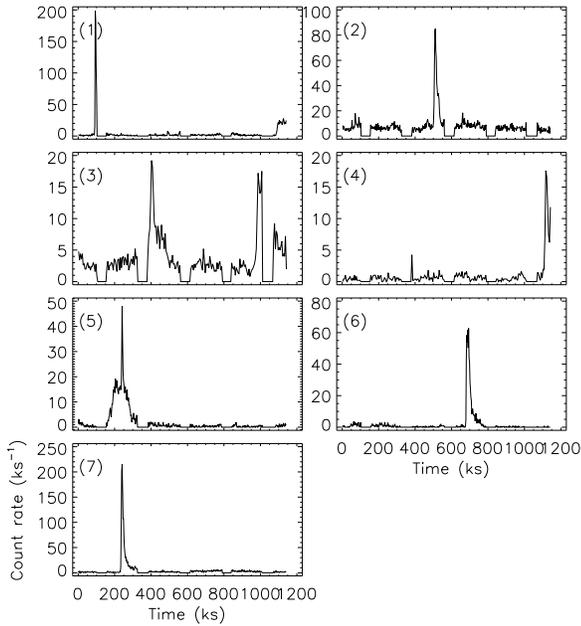}
 \caption{Light curves of 6.4~keV sources in 2.0--9.0~keV band. The abscissa is time
 from the start of the observation, while the ordinate is the detector count rate of
 each time bin. The light curves are binned with 2500 or 5000 s~bin$^{-1}$ according to
 their brightness.}\label{fg:f2}
\end{figure}

The light curves of the seven 6.4~keV sources are given in
Figure~\ref{fg:f2}. Time-sliced spectroscopy of these seven sources is not feasible
because of the paucity of the signal. We also checked the previous two short ACIS
observations of the ONC \citep{feigelson02} and found none of these sources have enough
counts for a similar analysis.

\subsection{Stellar Counterparts of the Control Sample Sources}
Most of the 1616 COUP sources, thus most of the control sample sources, are YSOs that
belong to Orion Nebula population. However, a significant contamination comes from
background AGNs. We estimate no more than $\approx$~6 AGNs in the control sample based
on the following calculation. The threshold of the control sample of 100 counts in
the 6.0--9.0~keV band is converted to an X-ray flux of
$\sim$~3.3$\times$10$^{-14}$~ergs~s~cm$^{-2}$ (2.0--7.0~keV) assuming a power-law
spectrum of an index typical of AGNs (--1.7) absorbed with a column density of
10$^{22}$~cm$^{-2}$, as a rough estimate of the averaged value in the COUP field. Using
\begin{equation}
 N(>S) = 1200 \left(\frac{S}{2 \times 10^{-15}~\rm{ergs~s^{-1}~cm^{-2}}}\right)^{-1.0}
\end{equation}
from \citet{giacconi01} where $N(>S)$ is the number of background X-ray sources per
square degree brighter than $S$~ergs~s~cm$^{-2}$, no more than $\approx$~6 AGNs are
expected above the 100 count threshold. We believe four of these AGN sources can be
identified: COUP sequence numbers 8, 723, 1053, and 1304 \citep{getman04}. They exhibit
no flare-like flux variations, show no spectral lines, and have best-fit \kt\ above
15~keV. Unphysically high temperature values are often introduced by fitting a flat
power-law spectrum with a thermal model. We are thus confident that the remaining 123
sources are Orion population members and consider them to be the control sample hereafter.

The seven sources with the 6.4~keV feature clearly have YSO properties: 6.7~keV line
emission arising from the K$\alpha$ line of Fe~$_{\rm{XXV}}$ (Fig.~\ref{fg:f1}), light
curves showing high amplitude flux variation typical of YSO flares (Fig.~\ref{fg:f2}),
and NIR identifications except for No. 4 (Table~\ref{tb:t2}). The lack of a NIR
counterpart for No. 4 is probably due to an exceptionally high column density exceeding
3$\times$10$^{23}$~cm$^{-2}$ (Table~\ref{tb:t1}).

\begin{deluxetable}{ccrrrcrc}
 \tabletypesize{\scriptsize}
 \tablecaption{NIR Colors of 6.4~keV Sources\label{tb:t2}}
 \tablecolumns{8}
 \tablewidth{0pt}
 \tablehead{
 \colhead{No.} &
 \colhead{\textit{JHK$_{\rm{s}}$}\tablenotemark{a}} &
 \colhead{\textit{J}} &
 \colhead{\textit{H}} &
 \colhead{\textit{K$_{\rm{s}}$}} &
 \colhead{\textit{L$^{\prime}$}\tablenotemark{b}} &
 \colhead{\textit{L$^{\prime}$}} &
 \colhead{TPSC\tablenotemark{c}} \\
 \colhead{} &
 \colhead{Counterparts} &
 \colhead{(mag)} &
 \colhead{(mag)} &
 \colhead{(mag)} &
 \colhead{Counterparts} &
 \colhead{(mag)} &
 \colhead{} 
 }
 \startdata
 1 & 05350920$-$0530585$^{\dagger}$ & 12.5 & 10.6 &  9.4 & \nodata & \nodata & \nodata \\
 2 & 387 & 11.1 &  9.4 &  8.3 & 1007 & 6.7 & 18 \\
 3 & 428 & 14.9 & 11.3 &  8.6 & 598 & 5.7 & 43 \\
 4 & \nodata & \nodata & \nodata & \nodata & \nodata & \nodata & \nodata \\
 5 & 446 & 12.6 & 11.4 & 10.8 & 175$^{\ddagger}$ & 10.5 & \nodata\\
 6 & 801 & 16.0 & 12.4 &  9.9 &  42 & 7.4 & 41\\
 7 & 05351965$-$0513264$^{\dagger}$ & 14.0 & 11.5 & 10.2 & \nodata & \nodata & \nodata\\
 \enddata
 \tablenotetext{a}{\textit{J}, \textit{H}, and \textit{K$_{\rm{s}}$} magnitudes are
 either from the 2MASS all-sky survey catalog ($\dagger$) or McCaughrean et al. (in
 prep.).}
 \tablenotetext{b}{\textit{L$^{\prime}$} magnitudes are either from \citet{muench02} or \citet[$\ddagger$]{lada04}. Nos. 1 and 7 are located outside the \textit{L$^{\prime}$}-band surveys.}
 \tablenotetext{c}{Numbers of the protostar candidate sources listed in \citet{lada00}.}
\end{deluxetable}

\section{DISCUSSION}
We have now established that seven of 123 hard and bright COUP stellar X-ray sources show
6.4~keV line emission. All of them exhibit powerful X-ray flares and thermal plasma line
emission typical of YSOs. Most of them also have NIR photospheric counterparts. We now
proceed to make inferences about these sources concerning the nature of these stars and
the origin of the fluorescent line emission.

\subsection{Location of the Fluorescing Material}
The observed EW values give us a constraint on the geometry of the continuum source and the
reflector that gives rise to the fluorescence. The following calculation demonstrates the
improbability that the fluorescent line arises in circumstellar or interstellar matter
along the line of sight of the continuum X-rays.

EW values of the K$\alpha$ fluorescent line is related to the incident hot continuum and
ambient cold material as \citep{liedahl98}
\begin{eqnarray}
 EW_{\rm{K\alpha}}
  &=&\frac{L_{\rm{K\alpha}}}{I(E_{\rm{K\alpha}})}\nonumber\\
  &=&\left(\frac{\Delta \Omega}{4\pi}\right)Y_{\rm{K\alpha}}\frac{E_{\rm{K\alpha}}}{I(E_{\rm{K\alpha}})}\int{n_{\rm{Fe}}(s)ds}\nonumber\\
 &\times&\int_{\chi}^{\infty}{\frac{I(E^{\prime})}{E^{\prime}}\sigma_{\rm{Fe}}(E^{\prime})dE^{\prime}}.\label{eq:e1}
\end{eqnarray}

\noindent
Here, we suppose that the incident X-ray with a spectrum of $I(E)$ is photoelectrically
absorbed by a reflector that subtends the angle $\Delta \Omega$ seen from the continuum
source. The reflector, which is assumed to be optically thin to hard X-rays, has an iron
density of $n_{\rm{Fe}}(s)$ along the line of ray ($s$) with the photoelectric cross
section of $\sigma_{\rm{Fe}}(E)$. The energy integral begins at the iron K edge energy
$\chi$. The K$\alpha$ fluorescent line with the luminosity $L_{\rm{K\alpha}}$ is
subsequently emitted at the energy $E_{\rm{K\alpha}}$ with a fluorescence yield of
$Y_{\rm{K\alpha}}$. Assuming a thermal bremsstrahlung spectrum for the continuum source;
$I(E) \propto \exp{(-E/k_{\rm{B}}T)}$ and
$\sigma_{\rm{Fe}}(E)=(E/\chi)^{-3}\sigma_{\rm{Fe}}(\chi)$, we obtain
\begin{eqnarray}
 EW_{\rm{K\alpha}}
  &=&\left(\frac{\Delta \Omega}{4\pi}\right)Y_{\rm{K\alpha}}E_{\rm{K\alpha}}N_{\rm{H}}^{\prime}Z_{\rm{Fe}}\sigma_{\rm{Fe}}(\chi)\nonumber\\
 &\times&\exp{\left(\frac{E_{\rm{K\alpha}}}{k_{\rm{B}}T}\right)}\left(\frac{k_{\rm{B}}T}{\chi}\right)^{-3}\nonumber\\
  &\times&\int^{\infty}_{\chi/k_{\rm{B}}T}\exp{(-x)}x^{-4}dx.
\end{eqnarray}

The integral along $s$ is converted to an equivalent hydrogen column density
($N_{\rm{H}}^{\prime}=\int n_{\rm{H}}(s)ds$) using
$n_{\rm{Fe}}(s)=Z_{\rm{Fe}}n_{\rm{H}}(s)$, where $n_{\rm{H}}(s)$ is the hydrogen density
and $Z_{\rm{Fe}}$ is the elemental abundance of iron relative to hydrogen in the
reflector. Substituting $Y_{\rm{K\alpha}}=0.34$ \citep{kortright01},
$E_{K\alpha}=6.40$~keV, $\chi=7.11$~keV,
$\sigma_{\rm{Fe}}(\chi)=2\times10^{-20}$~cm$^{-2}$ \citep{gullikson01}, and
$Z_{\rm{Fe}}=3\times10^{-5}$ \citep{daeppen00}, $EW_{\rm{K\alpha}}$ is evaluated as
\begin{equation}
 EW_{\rm{K\alpha}} = \alpha \left(\frac{\Delta \Omega}{4\pi}\right) \left(\frac{N_{\rm{H}}^{\prime}}{10^{22}~\rm{cm}^{-2}}\right) [\rm{eV}],\label{eq:e3}
\end{equation}
where $\alpha$ is a constant that has a weak dependence on \kt. Between 3--10~keV in
which our 6.4~keV sample sources are distributed (Table~\ref{tb:t1}), $\alpha$ has an
approximate value of $\sim$~2.5 (1.8--3.1). The value is slightly smaller than the one
($\sim$~3.3) calculated for Seyfert galaxies with a power-law incident spectrum of
$I(E)\propto(E/\chi)^{-0.7}$ (e.g., \citealt{krolik87}), reflecting the exponential
cut-off of thermal spectra at the hard end.

If the fluorescence is caused by circumstellar or interstellar matter along the line of
sight of the continuum X-rays, the $N_{\rm{H}}^{\prime}$ values would be the same with
$N_{\rm{H}}$ derived from the X-ray spectral fits (Table~\ref{tb:t1}). This can
account for only a few percent of the observed EW values as $\Delta \Omega/4\pi \le 1$
(eq. [\ref{eq:e3}]). A geometry is thus required in which the photoionizing X-rays suffer
a localized absorption much larger than that in the line of sight. This is consistent with
the result by \citet{imanishi01} concerning YLW\,16A, where they set an upper limit of
$\sim$~20~AU upon the distance from the protostar to the reflector, based on the lack of
detectable time lag between the flare start and the appearance of the fluorescent
emission.

\subsection{Absorption \& NIR Colors of the Fluorescent X-ray Sources}
Two conceivable geometries to realize localized reflection are: (1) the reflection by
the stellar photosphere like in the Sun; and (2) the reflection by a circumstellar
structure like a disk. Our examination of the X-ray absorption and the NIR colors of the
6.4~keV sources in comparison with the control sample points to a circumstellar origin.

\begin{figure}
 \figurenum{3}
 \epsscale{1.0}
 \plotone{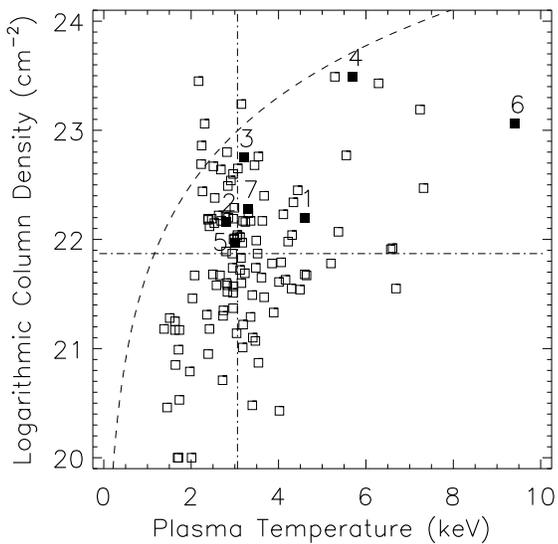}
 \caption{Scatter plot of plasma temperature and column density of the control
 sample (\textit{squares}). Sources with the 6.4~keV feature are marked filled with
 their source numbers. The dashed-and-dotted lines indicate the median values of
 \nh\ and \kt\ of the control sample. The dotted curve is the column density to have
 the opacity $\tau=1$ to the X-rays at a given temperature, which roughly indicates the
 observational bias.}\label{fg:f3}
\end{figure}

Figure~\ref{fg:f3} shows a scatter plot between \nh\ and \kt. The median value of \nh\
amongst the 123 control sample sources is 10$^{21.9}$~cm$^{-2}$. All of the 6.4~keV
sources show larger absorption than this. The preference, which has a null hypothesis
probability of $\sim$~0.8\%, is not expected if the reflection occurs at photosphere. It
would be likely, on the other hand, when the reflection takes place at disks, because
YSOs with disks are younger in their evolution than those without disks and tend to have
denser circumstellar matter yielding larger \nh.

\begin{figure}
 \figurenum{4}
 \epsscale{1.0}
 \plotone{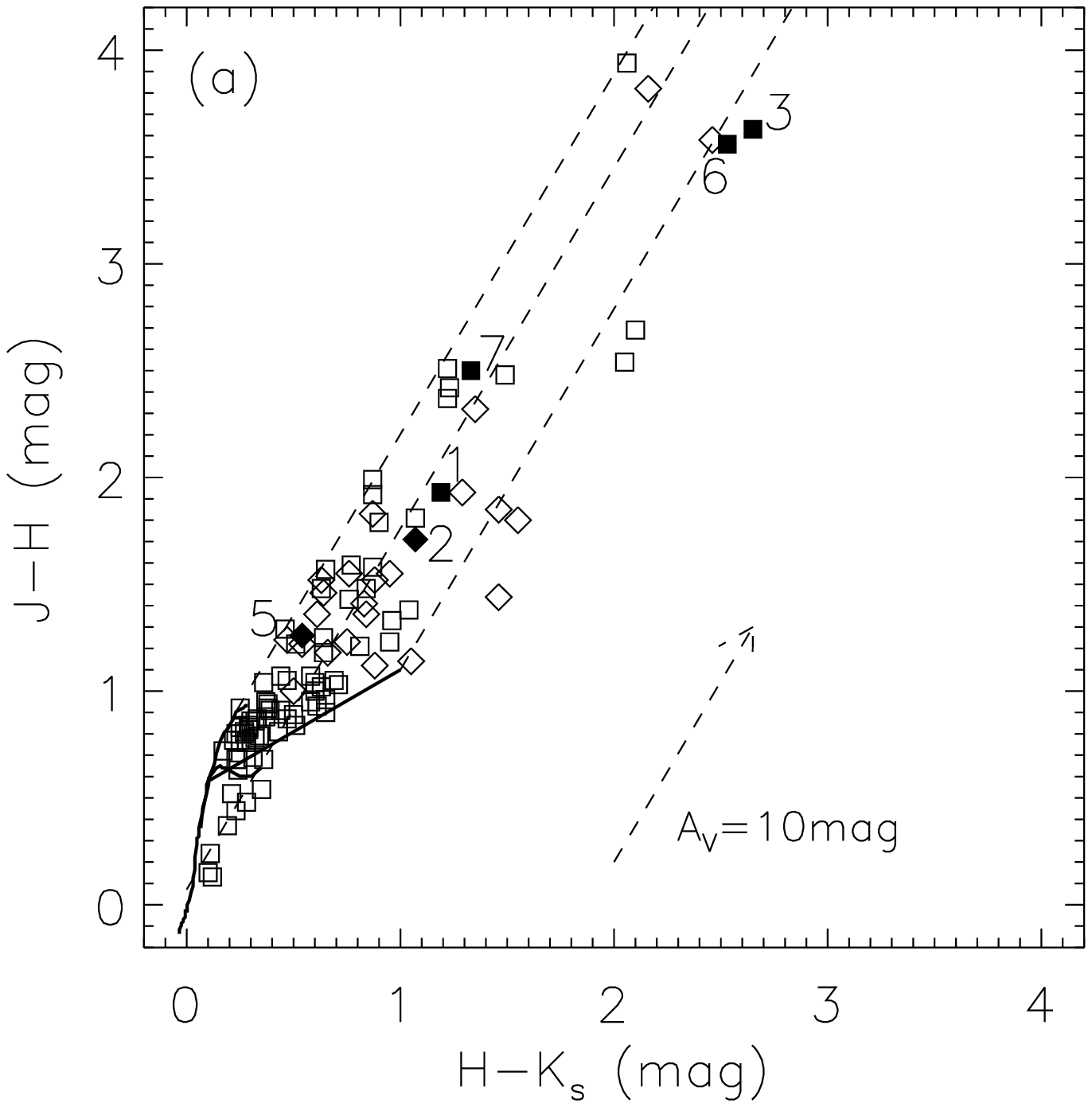}
 \plotone{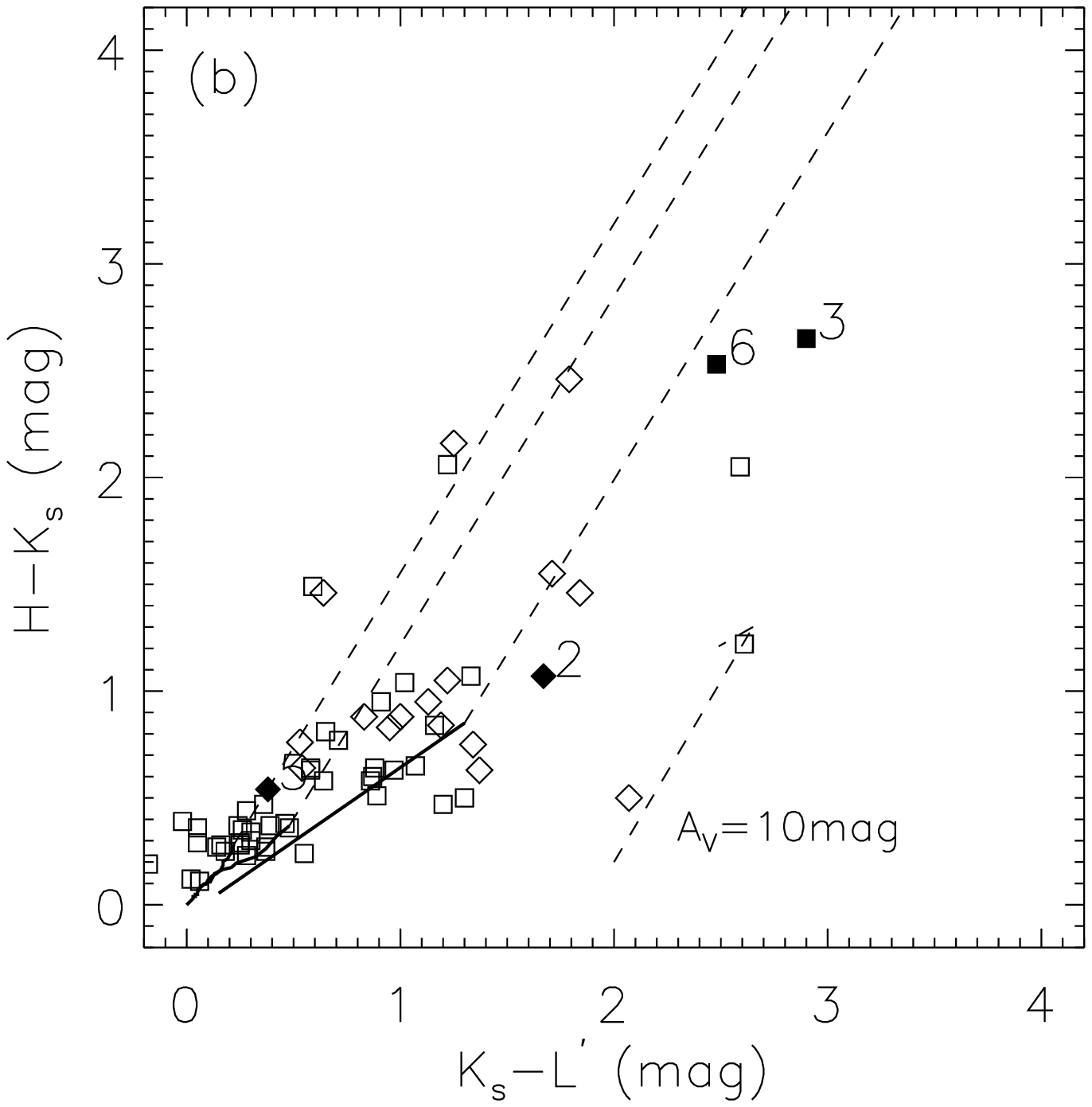}
 \caption{(\textit{J}--\textit{H})/(\textit{H}--\textit{K$_{\rm{s}}$}) and
 (\textit{H}--\textit{K$_{\rm{s}}$})/(\textit{K$_{\rm{s}}$}--\textit{L$^{\prime}$})
 diagrams (\textit{a} and \textit{b}) of the control sample sources. Squares are for
 sources with larger \kt\ and \nh\ than the medium values, while diamonds are for
 remainders. The 6.4~keV sources are marked filled with a label of their source
 numbers. The reddening lines are shown with dashed lines from the intrinsic colors of
 dwarfs and giants (\textit{thick solid curves}; \citealt{tokunaga00}) and the classical
 T Tauri star locus (\textit{thick solid lines}; \citealt{meyer97}). No conversion of the
 color systems is made to the data.}\label{fg:f4}
\end{figure}

Using \textit{J}-, \textit{H}-, \textit{K$_{\rm s}$}-, and \textit{L$^{\prime}$}-band
magnitudes of the NIR counterparts for the control sample sources, we show color-color
diagrams in Figure~\ref{fg:f4}. The NIR identifications and magnitudes follow
\citet{getman04}. Among 123 YSO samples, 109 have significant \textit{J}-, \textit{H}-,
and \textit{K$_{\rm{s}}$}-band detections and 66 additionally have significant
\textit{L$^{\prime}$}-band detections to be plotted on the
diagrams. The (\textit{J}--\textit{H})/(\textit{H}--\textit{K$_{\rm{s}}$}) and
(\textit{H}--\textit{K$_{\rm{s}}$})/(\textit{K$_{\rm{s}}$}--\textit{L$^{\prime}$})
diagrams (Fig.~\ref{fg:f4} \textit{a} and \textit{b}) complement with each other, where
the former covers fainter sources using shorter wavelengths data while the latter is
more sensitive to cooler circumstellar matter by using longer wavelengths data. At least
four 6.4~keV sources (Nos. 1, 2, 3, and 6) out of six with NIR counterparts lie outside
the region of reddened dwarfs and giants indicating the presence of NIR-emitting inner
disks. The preference to have NIR excess emission is consistent the idea that
fluorescent reflection occurs at disks.

\subsection{Fluorescence and Flares}
All the 123 control sample sources have more Bayesian block segments than two,
indicating flux variation. Most of them exhibit variation typical to flares with a fast
rise and slow decay profile. All of the seven sources with iron fluorescence have large
amplitude flares (Fig.~\ref{fg:f2}).

Our sample is heavily biased for flare sources. Therefore, we do not conclude that a flare
is a required factor for fluorescence, despite the facts that all fluorescence sources
exhibit flares and that we did not favor for flare sources in identifying the
fluorescence line by fitting time-integrated spectra.

\begin{figure}[h]
 \figurenum{5}
 \epsscale{1.0}
 \plotone{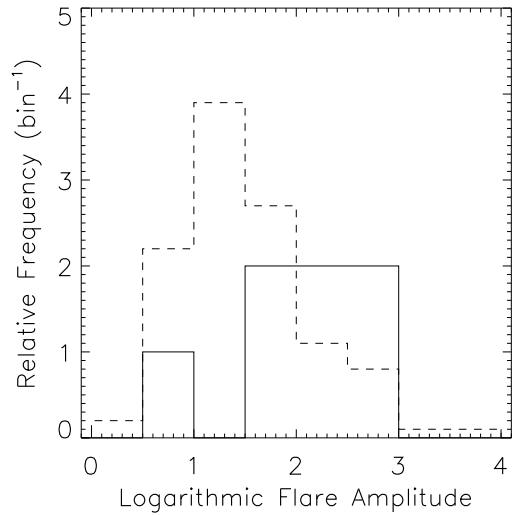}
 \caption{Histograms of flare amplutude for the 6.4 keV sources (\textit{solid}) and the
 control sample sources (\textit{dashed}). The frequency of the control sample sources
 is reduced by a factor of 10 to facilitate comparison.}\label{fg:f5}
\end{figure}

The flare amplitude, however, appears to be related to the
fluorescence. Figure~\ref{fg:f5} shows the histograms of flare amplitude of the control
sample and 6.4~keV sources, where we define the flare amplitude as the ratio between the
maximum and the minimum count rates of all Bayesian block segments. Six out of seven
6.4~keV sources have larger amplitudes than the median value ($\sim$10$^{1.42}$) among
the control sample, making the null hypothesis probability of this preference to be
$\sim$6\%. Large amplitude flares may have a preferable geometry for fluorescence to
occur, which should be examined by future studies.

\subsection{Constraints on the Geometry of the Fluorescent Sources}
We consider that the lack of 6.4~keV detections from the other 116 sources is not
completely attributable to a statistical inadequacy of the data. About 14\% of the 116
YSO samples without a 6.4~keV detection show: (1) \nh\ and \kt\ larger than the median
values and (2) significant NIR excess emission (squares located right to the middle
reddening lines of Fig.~\ref{fg:f4} \textit{a} and \textit{b}). Most of them have more
counts in the 6--9~keV band than the weakest of the seven 6.4~keV sources (107 counts
for No.\,5; Table~\ref{tb:t1}), which is close to the threshold for the control
sample. The absence of the fluorescence line thus appears to be a physical not
data-dependent statistical characteristic.

One idea is the disk orientation, if the fluorescence occurs at disks illuminated
from atop (the lamp post geometry). The geometry is considered a likely explanation for
the iron fluorescent emission from the accretion disks in Seyfert galaxies and X-ray
binaries \citep{nayakshin01}. Face-on inclination is more favorable than edge-on
inclination for the fluorescence X-rays to escape. \citet{george91} presented a relation
between the expected EW of the fluorescent line and the disk inclination with a
radiative transfer treatment. The calculation assumed a configuration where a continuum
source is at a height of 0.01$r$ above the center of a disk with a radius of $r$. The EW
value decreases drastically for large inclination angles relative to the normal of the
disk \citep[see][Fig.~14]{george91}. It may be the case that the bright sources without
a 6.4~keV detection have unfavorable inclination angles seen from us.

A simple consideration on the observed EW values also supports this picture. The iron
fluorescence can come out from any depth from the surface of disks corresponding to
different $N_{\rm{H}}^{\prime}$ values, but we speculate that the observed fluorescence
is dominated by the light reflected at $N_{\rm{H}}^{\prime} \sim 10^{24}$~cm$^{-2}$,
where 6.4~keV X-rays suffer photoelectric absorption of the opacity $\tau \lesssim
1$. We would expect a smaller $EW_{\rm{Fe}}$ if $N_{\rm{H}}^{\prime}$ is smaller than
$\sim 10^{24}$~cm$^{-2}$ by equation (\ref{eq:e3}). The claim is also the case if
$N_{\rm{H}}^{\prime}$ is larger, because the emitted 6.4~keV X-rays would be severely
attenuated through the reflector \citep{makishima86}. By substituting
$N_{\rm{H}}^{\prime} \sim 10^{24}$~cm$^{-2}$ and $EW_{\rm{Fe}} \sim 140$~eV, which is
the mean value of all seven sources (Table~\ref{tb:t1}), we can derive $(\Delta
\Omega/4\pi) \sim 0.5$. This is consistent with the assumption that the disk is
illuminated in a lamp post configuration to realize the subtended angle, and the disk is
oriented face-on so that the fluorescent X-rays suffer no extra attenuation through the
disk.

The detection of the 6.4~keV line has an astrophysical implication on X-ray ionization
of disks. Many studies (e.g., \citealt{glassgold99}) discussed that X-ray
photoionization, in addition to cosmic-ray collision ionization, can be a dominant
ionization source for circumstellar material around forming stars, which is an important
process to couple the gas with the ambient magnetic field. Our detection of the
recombination line at 6.4~keV is a direct observational proof that the X-ray
photoionization does take place at disks.

Finally, we note that the reflection may occur in different disk geometries such as
X-ray irradiation at the inner edge of the disk, or at other circumstellar structures
such as funnel flow, jet, or wind columns; these structures are associated with sources
at an early stage of YSO evolution. So far, we have no observational argument to
distinguish among these possibilities. The measurement of the delay time from the onset
of flares to the appearance of the 6.4~keV line, as was tried by \citet{imanishi01} for
an upper limit value, will bring a key observational basis by future X-ray experiments
like \textit{Constellation-X}.

\acknowledgments
The authors express gratitude for Konstantin Getman for his help in data reduction, and
Salvatore Sciortino and Koji Mori for improving the manuscript. COUP is supported by
\textit{Chandra} guest observer grant SAO GO3-4009A (E. D. Feigelson, PI) and the ACIS
Team contract NAS8-38252 (G. P. Garmire, PI). M.\,T. acknowledges financial supported by
the Japan Society for the Promotion of Science.


\end{document}